\documentclass[twocolumn,showpacs,preprintnumbers,amsmath,amssymb]{revtex4}

\usepackage{graphicx}% Include figure files
\usepackage{dcolumn}% Align table columns on decimal point
\usepackage{bm}% bold math

\begin{document}

\title {Magnetization dynamics in the inertial regime: nutation
predicted at short time scales} \author{M.-C. Ciornei} \author{J.-E.
Wegrowe} \affiliation{Ecole Polytechnique, LSI, CNRS and
CEA/DSM/IRAMIS, Palaiseau F-91128, France.} \author{J. M. Rub\'i}
\affiliation{Departement de Fisica Fonamental, Universitat de
Barcelona, Diagonal 647, Barcelona 08028, Spain.}

\date{\today}

\begin{abstract}
The dynamical equation of the magnetization has been reconsidered with
enlarging the phase space of the ferromagnetic degrees of freedom to
the angular momentum.  The generalized Landau-Lifshitz-Gilbert
equation that includes inertial terms, and the corresponding
Fokker-Planck equation, are then derived in the framework of
mesoscopic non-equilibrium thermodynamics theory.  A typical
relaxation time $\tau$ is introduced describing the relaxation of the
magnetization acceleration from the inertial regime towards the
precession regime defined by a constant Larmor frequency.  For time
scales larger than $\tau$, the usual Gilbert equation is recovered. 
For time scales below $\tau$, nutation and related inertial effects
are predicted.  The inertial regime offers new opportunities for the
implementation of ultrafast magnetization switching in magnetic
devices.
\end{abstract}

\pacs{75.78.jp, 05.70.Ln  \hfill}

\maketitle

%\subsubsection{Introduction}
%\paragraph{\textbf{Introduction}}

In 1935 Landau and Lifshitz proposed an equation for the dynamics of
the magnetization $\mathbf{M} $ (of constant modulus), composed of a
precession term $\mathbf{M} \times \mathbf{H}$ and a longitudinal
relaxation term $\mathbf{M} \times (\mathbf{M} \times \mathbf{H})$
that drives the magnetization towards equilibrium along the magnetic
field $\mathbf{H}$ \cite{Landau}.  Two decades later T. L. Gilbert
derived the equation that bears his name in which the relaxation
towards equilibrium is described by a damping term $\eta$
\cite{Gilbert} through the dynamic equation $d\mathbf{M} /dt = \gamma
\mathbf{M} \times (\mathbf{H} - \eta d\mathbf{M} /dt)$, with $\gamma$
as the gyromagnetic ratio.  The two equations (Landau-Lifshitz and
Gilbert) are mathematically equivalent.

The range of validity of the Landau-Lifshitz-Gilbert (LLG) equation was established one decade later by W. F. Brown, with a description
of a magnetic moment coupled to a heat bath ("thermal fluctuations of a
single-domain particle", 1963 \cite{Brown}).  The magnetic moment is treated as a Brownian particle described by the slow degrees of freedom ($10^{-9}\,s$), the angles $\{\theta,\phi\}$. The remaining degrees of freedom of the system relax in a much shorter time scale ($ < 10^{-12}\,s$). The time scale separation between the rapidly relaxing environmental degrees of freedom and the slow magnetic degrees of freedom allows the coupling between the magnetization and the environment to be reduced to a single phenomenological damping parameter $\eta$, whatever the complexity of the microscopic relaxation involved \cite{Fick,Coffey}.
 
However, important experimental advances towards very
short time-resolved response of the magnetization (sub-picoseconds
resolution, i.e. below the limit proposed by Brown) have been reported in
the last decade \cite{Rapid}.  In parallel, industrial needs for very
fast memory storage technologies are approaching the limits imposed by
the precessional switching \cite{Tudosa}.  In these experiments and in
the corresponding applications, time scale separation between the
conserved degrees of freedom $\{\theta,\phi\}$ and the other degrees
of freedom, assumed by Brown \cite{Brown}, finds its limit.

The purpose of this Letter is to investigate the dynamics of the
magnetization beyond this limit by extending the phase space to
additional degrees of freedom expected to be also out-of-equilibrium
at short time scales \cite{Fick, RubiInertial}.  According to the
well-known gyromagnetic relation \cite{Einstein}, the next relevant
degree of freedom of the ferromagnetic
system (beyond the coordinates of position; i.e. the angles
$\{\theta,\phi \}$) is the angular momentum $\mathbf{L}$.  As will be
shown below, the consequence of considering also the conservation of
the angular momentum is that inertial terms, i.e. acceleration terms
proportional to $d^{2} \mathbf{M}/dt^{2}$, appear in the equation of
motion.  The existence of inertial terms in the dynamics of the
magnetization opens the way to deterministic ultrafast magnetization
switching strategies, beyond the limitations of the precessional
regime \cite{Kimel}.  We assume however that the microscopic
relaxation channels are inactive at the time considered here (e.g. by
choosing the adequate materials and excitations of the magnetization). 
Otherwise, a non-deterministic regime would take place
\cite{Tudosa,Bigot,PRB08}.

We derive below the generalized Gilbert equation and the corresponding
Fokker-Planck equation that includes the inertial effects for a
uniform magnetic moment.  The derivation is performed in the framework
of mesoscopic nonequilibrium thermodynamics (MNET) \cite{Rubi, Mazur,
DeGroot}, and is based on the expression of the conservation laws,
thermodynamic laws, and symmetry properties.

It is convenient to model the dynamics of a magnetic moment $\mathbf{M}=M_s\mathbf{e}$ (submitted to an applied magnetic field $\mathbf{H}=-\frac{1}{M_s}\frac{\partial V^F}{\partial \mathbf{e}}$ and coupled to a heat bath) with a statistical ensemble composed by non-interacting identical uniform magnetic moments found in the same given conditions (ergodic property). Here, $\mathbf{e}$, $M_s$ and $V^{F}$ are respectively the radial unit vector of angles $\{\theta,\phi\}$, the magnetization at saturation, and the ferromagnetic potential energy. The ensemble of magnetic moments of constant modulus $M_s$ defines a sphere surface $\Sigma$ and the number of magnetic moments oriented within $(\mathbf{e},\mathbf{e}+d\mathbf{e})$ defines the density $n(\mathbf{e})$ of magnetic moments over $\Sigma$. We have shown in previous works that associating two degrees of freedom $\{\theta,\phi\}$ to a magnetic moment is sufficient to derive both the Gilbert equation and the corresponding rotational Fokker-Planck equation from non-equilibrium thermodynamics principles only \cite{JPhys,PRB08}.
   
Extending the configuration space to the {\it magnetic angular momentum} $\mathbf{L}$, the space $\Sigma$ is extended from a two dimensions space, to {\it a priori} five dimensions space $\{\theta,\phi, \mathbf{L}\}$   \cite{Rque2}.
    A distribution function $f(\mathbf{e},\mathbf{L})$ of magnetic moments with the magnetization orientation within $(\mathbf{e},\mathbf{e}+d\mathbf{e})$ and the angular momentum within $(\mathbf{L},\mathbf{L}+d\mathbf{L})$ should then be defined, where $f$ is assumed to vanish for infinite values of $\mathbf{L}$ as: \(
\lim_{\mathbf{L}\rightarrow \pm\infty} f(\mathbf{e},\mathbf{L})=0 \).  
The angular momentum $\mathbf{L}$ associated to a magnetic moment is either changed by an applied torque $\mathbf{N} = \mathbf{M} \times \mathbf{H}$ as $\left (\frac{d\mathbf{L}}{dt} \right )_{s}= \mathbf{N}$, either by the interaction with the heat bath. When considering the statistical ensemble, the interaction with the bath is modeled through a phase space flux $\mathbf{J_L}$ (defined below)
 which vanishes for large values of $\mathbf{L}$: $\lim_{\mathbf{L}\rightarrow \pm \infty} \mathbf{J}_{L}=0$. 

 The kinetic energy expression that Gilbert associated to the magnetization \cite{Gilbert} is written as: $\mathcal{K}= \mathbf{L} \mathbf{L} : \bar{\bar{I}}^{-1}/2$, where the magnetic inertial tensor $\bar{\bar I}$, is related to the magnetic moment (and not to the inertia of matter). It is assumed that $\bar{\bar I}$ keeps the symmetry of the magnetic moment, i.e. is axial symmetric of symmetry axis $\mathbf{e}$: $\bar{\bar I}= I_1\! \left (\bar{\bar{U}} - \mathbf{e} \mathbf{e} \right) + I_{3}\mathbf{e} \mathbf{e}$, with $\bar{\bar U}$ the dyadic unit (where $I_1 = I_2$ and $I_3$ are the diagonal coefficients of the inertial tensor). 

In the space-fixed reference frame denoted by the subscript $s$, the conservation law for the number of axial symmetric moments $f(\mathbf{e},\mathbf{L})$ writes \cite{Condiff}:
    \begin{equation}
       \frac{\partial}{\partial t} f(\mathbf{e},\mathbf{L}_{s}) =
         - 
            \left \{ 
              \frac {\partial \, ( f  \mathbf{\dot{e}} ) }
                    {\partial \mathbf{e}} 
           \right \}_{\mathbf{L}_s} \,
         - 
           \mathbf{N}_{s} 
           \cdot
           \frac{\partial f} {\partial \mathbf{L}_{s}}
         - 
           \frac{\partial \mathbf{J_L}} {\partial \mathbf{L_{s}}}
     \label{Boltzm_cart}
    \end{equation}
where the derivatives with respect to the angles are made while holding the Cartesian components of $\mathbf{L}_s$ constant:
\begin{equation}
\!\!
 \left \{ 
              \frac {\partial  ( f  \mathbf{\dot{e}} ) }
                    {\partial \mathbf{e}} 
           \right \}_{\mathbf{L}_s}
=
  \frac{1}{\sin \theta}
  \left\{
    \frac{\partial 
      (f\sin\theta\,\dot{\theta}  )
         }
        {\partial \theta}
  \right\}_{\mathbf{L}_s} 
  +
   \left\{
    \frac{\partial (f\dot{\phi})}{\partial \phi}
  \right\}_{\mathbf{L}_s} 
\end{equation}

The density $n(\mathbf{e})$ of magnetic moments in the space $\Sigma$ is recovered by integrating over the angular momentum degree of freedom $n(\mathbf{e}){=} \int f(\mathbf{e},\mathbf{L})\,d^{3}\mathbf{L}$. The conservation law for the magnetic moments in the $\Sigma$ space is hence deduced from \eqref{Boltzm_cart}:

\begin{gather}
  \frac{\partial n}{\partial t}
 =
     \int \frac{\partial f}{\partial t}\, d^{3} \mathbf{L}_{s}
 =
  -\frac{\partial } 
        {\partial \mathbf{e}}
    \cdot
    \int f  \mathbf{\dot{e}} \, d^{3}\mathbf{L}_{s}
 = 
 - \frac{\partial (n \dot{\mathbf{e}})} {\partial \mathbf{e}} 
   \label{cons_part_positions} 
   \end{gather} 

 Beyond, the conservation law for the mean value of the magnetic angular momentum $\langle \mathbf{L}_{s} \rangle $ is also derived \cite{Condiff,Ciornei}:
   \begin{align}
	   \frac{\partial \, n \langle \mathbf{L}_{s} \rangle}{\partial t}
       &
       =
          \int
            \frac{\partial f}{\partial t}\, \mathbf{L}_{s}
            \,d^{3} \mathbf{L}_{s} \nonumber \\
      &
      \stackrel
         {\eqref{Boltzm_cart}
           \
         } { = }
       - 
         \frac{\partial}{\partial \mathbf{e}}
         \cdot
         \int
            f \mathbf{\dot{e}}\, \mathbf{L}_{s}
           \, d^{3} \mathbf{L}_{s}
       + 
         n \,  \mathbf{N}_s(\mathbf{e})
       +
         \int
         \mathbf{J_L} \,d^{3} \mathbf{L}_{s} \nonumber
 \\
% \end{align}
%\begin{align} 
  % \Rightarrow
  n \, \frac{d \langle \mathbf{L}_{s} \rangle}{d t}
     & 
    =
       -
         \frac{\partial}
              {\partial \mathbf{e}}
         \cdot
         \left(
          \mathbf{e} \times 
          \overline{\overline{P}}_s
         \right)
       + 
          n \, \mathbf{N}_s(\mathbf{e})
       +
         \int
           \mathbf{J_L}
           \,d^{3} \mathbf{L}_{s}
   \label{gen_dynamic_eq}
   \end{align}
	where the {\it magnetic pressure} tensor is defined as
	$\overline{\overline{P}}_s= \overline{\overline{I}}^{-1} \int 
	(\mathbf{L} \, - \, \langle \mathbf{L}_{s} \rangle)(\mathbf{L} \, - \, 
	\langle \mathbf{L}_{s} \rangle) \, f\, d^3\mathbf{L}_{s}$.

The conservation equation \eqref{gen_dynamic_eq} states that the rate of the average angular momentum $\langle \mathbf{L}_{s} \rangle $ is due to three contributions: an applied torque $\mathbf{N}_s$, an average interaction with the bath $\int \mathbf{J_{L}}d^{3} \mathbf{L}_{s}$ (i.e. damping), and a torque due to pressure (i.e. rotational diffusion).

  The expression for $\mathbf{J_L}$ is deduced from the entropy production expression $\sigma(\mathbf{e})$ \cite{Ciornei, Rubi, Mazur}. Defining the ferromagnetic chemical potential $\mu(\mathbf{e},\mathbf{L})$, the power $T \sigma(\mathbf{e})$ dissipated by the magnetic system is the product of the generalized flux by the generalized force:

  \begin{equation}
  T \sigma(\mathbf{e}) = - \int \mathbf{J_L} \cdot \frac{\partial \mu} {\partial \mathbf{L}_s}  \,d^{3}\mathbf{L}_s  
  \label{entropy_prod}
 \end{equation}

  where the chemical potential takes the canonical form \cite{DeGroot, Rubi}:

  \begin{equation}
    \mu(\mathbf{e},\mathbf{L}) = kT \, ln \left [f(\mathbf{L}, \mathbf{e}) \right ] + 
   \mathcal{K}(\mathbf{e},\mathbf{L}) + V^{F}(\mathbf{e})
   \label{PotChem}
  \end{equation}

  The application of the second law of thermodynamics, together with the local equilibrium hypothesis in the $(\mathbf{e},\mathbf{L})$ space, lead us to the introduce the Onsager matrix $\overline{\overline{\mathcal{L}}}$ such that:
  $\mathbf{J_L} = - \overline{\overline{\mathcal{L}}}\cdot \frac{\partial \mu}{\partial \mathbf{L}_s}$. As the Onsager coefficients are a reflection of the system's symmetry \cite{Mazur}, the relaxation tensor defined as $\overline{\overline{\tau}}^{-1}=\frac{1}{f}\overline{\overline{\mathcal{L}}}\,\overline{\overline{I}}^{-1}$ is also axial symmetric: $\overline{\overline{\tau}}^{-1}=\tau_1^{-1}(\overline{\overline{U}}-\mathbf{e}\mathbf{e})+\tau_3^{-1}\mathbf{e}\mathbf{e}$ (where $\tau_1 = \tau_2$ and $\tau_3$ are the diagonal coefficients), and is related to damping. Moreover, as $\mathbf{e}$ is an axis of symmetry for the ferromagnetic potential $V^F(\theta,\phi)$, the relaxation tensor $\overline{\overline{\tau}}^{-1}$ is not expected to have any components in the $\mathbf{e}$ direction \cite{Condiff}, leading to $\tau_3^{-1}=0$.

The dynamic equation \eqref{gen_dynamic_eq} can be rewritten: 
\begin{equation}
\frac{d \langle \mathbf{L_{s}} \rangle}{d t}
     = \mathbf{N_{s}}
         -  \overline{\overline{\tau}}^{-1}_s 
            \cdot
            \langle \mathbf{L_{s}} \rangle
         - \frac{1}{n} 
           \frac{\partial }
                {\partial \mathbf{e}}
           \cdot
           \left(
           \mathbf{e} \times 
           \overline{\overline{P}}_s
           \right)
         \label{dynamic_eq}
\end{equation}

As the inertial tensor $\overline{\overline{I}}$ and the relaxation tensor $\overline{\overline{\tau}}^{-1}$ are time
independent in the rotating frame (or magnetization frame), a simpler expression of Eq. \eqref{dynamic_eq} can be obtained in this frame. After introducing the average angular velocity $\mathbf{\Omega}$ such that $\langle\mathbf{L}\rangle= \overline{\overline{I}}
\cdot\mathbf{\Omega}$, Eq. \eqref{dynamic_eq} rewrites as:

\begin{equation}
\frac{d \mathbf{\Omega}_{r}}{d t}
     = \overline{\overline{I}}^{-1}_{r} 
       \cdot  
       \left[
         \mathbf{N_{r}} 
        -
         \frac{1}{n}
         \frac{\partial }
              {\partial \mathbf{e}} 
           \cdot
           \left(
            \mathbf{e} \times   
             \overline{\overline{P}}
           \right)
       \right]
      -
       \overline{\overline{\tau}}^{-1}_{rot} 
       \cdot \mathbf{\Omega}_{r}
   \label{dyn_eq_rotating_frame}
\end{equation}
The rotating frame is denoted by the subscript $r$ and  $\overline{\overline{\tau}}^{-1}_{rot} =
\overline{\overline{\tau}}^{-1}_r - \left ( \frac{I_{3}}{I_{1}} - 1
\right ) \Omega_{3}\, \mathbf{e} \times \overline{\overline{U}}$, or 

\begin{equation}
\overline{\overline{\tau}}_{rot}^{-1} = 
(\tau_1 \, \alpha^{*})^{-1} \,  \left( \begin{array}{ccc}
  \alpha^{*} &  1 & 0 \\
                        - 1 & \alpha^{*} & 0 \\
						0 &  0 & 0 \\
\end{array} \right)
\label{MatrixBeta}
\end{equation}

where $\alpha^{*} = \alpha \,  (I_3/I_1 - 1 )^{-1}$ with $\alpha = (\Omega_3 \,  \tau_1)^{-1}$.

The three components of Eq. \eqref{dyn_eq_rotating_frame} read:
\begin{equation}
\!\!\!\!
\left\{
        \begin{aligned}
 \dot{\Omega}_{1} & = 
  - \frac{\Omega_{1}}{\tau_1} 
    - 
      \left ( 
        \frac{I_{3}}{I_1} - 1 
      \right)
      \! \Omega_{3} \Omega_{2} 
    - \frac{M_{s}H_{2}}{I_1}  
    - \!
      \left[
        \frac{1}{I_1n} 
        \frac{\partial 
         (
          \mathbf{e}
          \times
          \overline{\overline{P}}
          )}
         {\partial \mathbf{e} } 
      \right]_{1}
 \\
 \dot{\Omega}_{2} & 
  =  - \frac{\Omega_{2}}{\tau_1}
    + 
      \left (
        \frac{I_{3}}{I_1} - 1 
      \right)
      \!\Omega_{3}\Omega_{1} 
    +
      \frac{ M_{s} H_{1} } {I_1} 
    -\!
      \left[
        \frac{1}{I_1n} 
        \frac{\partial 
         (
          \mathbf{e}
          \times
          \overline{\overline{P}}
          )}
         {\partial \mathbf{e} } 
      \right]_{2} \\
\dot{\Omega}_{3} &= - \tau_3^{-1} \Omega_3 = 0
\end{aligned}
\right.
\label{OmegaCoord}
\end{equation}

Since the quantity $ L_{3} = I_3 \Omega_{3}$ is a constant of motion, the well-known gyromagnetic ratio $\gamma$ can be defined as the ratio of the magnetization at saturation $M_s$ by the axial angular momentum $\gamma =\frac{M_{s}}{\langle L_{3} \rangle }$ \cite{Einstein,Gilbert}. 

Also, the averaged dynamic Eq. \eqref{OmegaCoord} introduces a characteristic time scale $\tau_1$, which separates the behavior of the magnetic system of particles in two regimes: the diffusion regime or the long time scale limit $t\gg\tau_1$, and the inertial regime or the short time scale limit $t\ll\tau_1$.  
Since the modulus of the magnetization  $\mathbf{M}$ is conserved, the relation $\frac{d \mathbf{M}}{dt} = \mathbf{\Omega} \times \mathbf{M}$ holds. 
 Cross-multiplying by $\mathbf{M}$ and using the above definition of
 $\gamma$ leads to the identity $ \mathbf{\Omega} =
 \frac{\mathbf{M}}{M_{s}^{2}} \times \frac{d\mathbf{M}}{dt} +
 \frac{\mathbf{M}}{I_{3} \gamma}$.  

In a diffusive regime, i.e. for $t\gg\tau_1$, the inertial terms $\frac{d\Omega_1}{dt}$ and $\frac{d\Omega_2}{dt}$ are negligible with respect to $\frac{\Omega_1}{\tau_1}$ and $\frac{\Omega_2}{\tau_1}$. Eq. \eqref{dyn_eq_rotating_frame} then rewrites as the Gilbert equation with an inertial correction performed on the previously defined gyromagnetic coefficient  $\gamma^{*} = \frac{\gamma}{1-I_1/I_{3}}$ :
  \begin{equation}
    \frac{d\mathbf{{M}}}{dt}
    =
   \gamma^{*}\,
    \mathbf{M}
    \times
    \left(
      \mathbf{H}_{eff}
   -
      \eta \frac{d\mathbf{{M}}}{dt}
    \right)
   \label{Gilbert_diff} 
  \end{equation}

  The Gilbert damping coefficient $\eta$ is now defined as: $\eta=\frac{I_1}{\tau_1M_s^2}$ (so that $\alpha^{*} = \gamma^{*} \eta M_{s} $ is the corresponding dimensionless coefficient), and $\mathbf{H}_{eff}$ is an effective field that includes the diffusion term.

At the diffusive limit, the magnetic moments follow a distribution function $f(\mathbf{e},\mathbf{L})$ close to a Maxwellian centered on the average angular momentum $\langle\mathbf{L}\rangle$ \cite{RubiInertial}. This leads to a diagonal form for the pressure tensor: $\overline{\overline{P}}=nkT/\overline{\overline{U}}$ and $\mathbf{H}_{eff} = \mathbf{H} - \frac{kT}{n}\frac{1}{M_s}\frac{\partial n}{\partial  \mathbf{e}} $ \cite{JPhys, PRB08}.

 Eq. (\ref{Gilbert_diff}) contains the density $n(\mathbf{e)}$ so that the equation is not closed. However, inserting Eq. (\ref{Gilbert_diff}) into the conservation law \eqref{cons_part_positions} leads to the rotational Fokker-Planck equation of $n(\mathbf{e)}$, derived by Brown \cite{Brown}:  $\frac{\partial n}{\partial t} = \frac{\partial  \left( n\mathbf{e} \times \mathbf{\Omega} \right) }{\partial \mathbf{e}}$

   For short enough time scales $t \approx \tau_1$, the inertial terms cannot be neglected and the Gilbert approximation is no longer valid. The dynamic equation \eqref{OmegaCoord} takes the following generalized form:

   \begin{equation}
    \frac{d\mathbf{\mathbf{M}}}{dt}
   =
    \gamma \mathbf{M}
    \times
    \left[
      \mathbf{H}
     -
      \eta 
      \left(
        \frac{d\mathbf{M}}{dt}
       +
        \tau_1
        \frac{d^2\mathbf{M}}{dt^2} 
      \right)  
    \right] 
   -
    \frac{ \gamma}{n}
    \frac
     {\partial (\mathbf{M}\times\overline{\overline{P}}) }
     {\partial \mathbf{M}}
   \label{motion_angular}
   \end{equation}

The corresponding generalized rotational Fokker-Planck equation for the statistical distribution $f$ is obtained with replacing $\mathbf{J_L}$ by the Onsager relation derived earlier into the conservation law \eqref{Boltzm_cart} and rewriting the law in the rotating frame \cite{Ciornei}:

\begin{gather}
 \frac{\partial f_{(\mathbf{e},\mathbf{L}_r)}}{\partial  t}
 =
  \left\{
    \frac
       {\partial 
         (
           f\mathbf{e}\times
           \overline{\overline{I}}^{-1}_r \mathbf{L}_r
          )
       }
       {\partial \mathbf{e}}
  \right\}_{\mathbf{L}_r} +
 \nonumber \qquad\qquad\qquad\qquad  \\
 \qquad\qquad\qquad 
 \frac{\partial}{\partial \mathbf{L}_r}
  \cdot
  \left[
    f \overline{\overline{\tau}}_{rot}^{-1}
    \cdot
        \mathbf{L}_r
     -
      f \mathbf{N}_r + kT \overline{\overline{\tau}}_{r}^{-1} \overline{\overline{I}}_{r}
      \cdot
      \frac{\partial f}{\partial \mathbf{L}_r}
  \right]
\end{gather}

At short time scales $t \approx \tau_1$ and due to the inertial effect,
the usual precessional behavior is enriched by a {\it nutation} effect.  The simplest way to understand nutation is to imagine that the effective field is switched off suddenly with zero damping: the precession stops suddenly because the Larmore frequency  $\omega_{L} = \gamma^{*} \mathbf{H}$ drops to zero at the same time. However, in the absence of inertial terms, the magnetic moment also stops at this position within an arbitrarily short time scale. But if the kinetic energy is not zero (and this is the case for magnetomechanical measurements of the magnetization \cite{Einstein}), the movement cannot be stopped suddenly: the precession (around the magnetic field) stops but the magnetic moment starts to rotate around the angular momentum vector in order to conserve the energy: the precession is transformed into nutation.

\begin{figure}[h!t]
\includegraphics[scale=0.4] {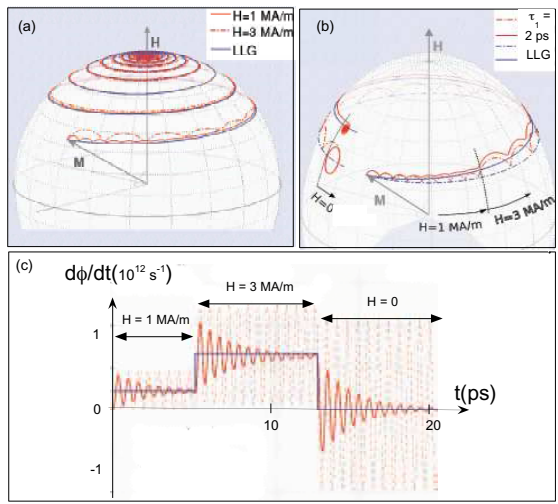}
 \caption{ Numerical resolution of Eq. (\ref{motion_angular}) with $\tau_1 = 2 ps$. (a) Trajectories at two different fields with $\vert \alpha \vert = 0.05$ (red) and curves deduced from the LLG equation (blue). (b) Trajectory of the magnetization with changing suddenly the effective fields from H=1 MA/m to H=3 MA/m and H=0, with damping (continuous line) and without (dotted line). (c) Time derivative of the azimuth angle $\phi$ plotted as a function of time for the trajectory of Fig. 1(b).}
\end{figure}

Fig. 1 shows the numerical resolution of Eq. (\ref{motion_angular}) (neglecting thermal fluctuations) with a field along z axis and for a parameters $\tau_1$ fixed to $2 ps$ with $\vert \alpha \vert= 0.05$. The trajectories are plotted on the sphere $\Sigma $. The usual trajectory deduced from the LLG equation ($\tau_1 \ll ps$) is also plotted for comparison. The motion of the magnetic moment displays the familiar curve due to Larmor precession, with superimposed loops generated by the nutation effect. Fig.1(b) shows a trajectory starting without initial velocity under an effective field of $1MA/m$, changed suddenly to $3MA/m$ and once again down to zero.  Four curves are represented, two for the Eq. (\ref{motion_angular}) with $\vert \alpha \vert= 0.05$ (red continuous line) and $\alpha = 0$ (red dashed line), and two for the usual LLG equation with and without damping (blue). At the end of the motion (left), the field is set to zero and the precession is destroyed, with the nutation effect shaping a circle (without damping) or a spiral (with damping). Note that the profile of the nutation loops depends on the initial conditions (the cusp presented in Fig.1 instead of loops is due to zero initial velocity). Fig. 1 (c) shows the time derivative of the angle $\phi$ as a function of time for the trajectory displayed in Fig. 1 (b). The horizontal lines represent the constant Larmor frequencies, and the oscillations are due to nutation (for $\vert \alpha \vert = 0.05$ and $\alpha = 0$).

In the case of two stable states separated by a potential barrier (e.g.  for magnetic memory units), efficient strategies based on the inertial mechanism can be perormed for ultrafast magnetization reversal. Such a strategy has already been implemented in the case of antiferromagnets \cite{Kimel}, in which inertial terms are present due to the energy stored by deformation in the magnetic domain walls. The inertia has been used to overcome an energy barrier after having push the magnetization with a very short optical impulsion.  The novelty of our results is that any kind of ferromagnets could in principle be used for ultra-short inertial magnetization switching.

In conclusion, we have shown that extending the phase space of the magnetization to the degrees of freedom of the magnetic angular momentum leads to considere a generalized Landau-Lifshitz-Gilbert equation that contains inertial terms. This extension is justified by the well-known gyromagnetic relation that relates the magnetization to the angular momentum. It is predicted that inertial effects should be observed at short enough time scales (typicaly below the picosecond), e.g.  by measuring nutation loops superimposed to the usual precession motion of a magnetic moment.  The inertial regime at short time scales would also offer possibilities for new experiments and devices based on ultrafast magnetization switching.

%\bibliography{bibl}

\end{document}